\documentclass[12pt]{article}

\usepackage{jheppub}
\usepackage{amssymb, amsmath,amsthm}
\usepackage{amsfonts}
\usepackage{dsfont}
\usepackage{graphicx}
\usepackage{calc,ifthen,setspace,amscd}
\usepackage{dsfont}
\usepackage{bbm}

\usepackage{xcolor}
\usepackage{esvect}

\usepackage{subcaption}
\usepackage{orcidlink}
\usepackage{cleveref}

\crefformat{equation}{(#2#1#3)}
\crefrangeformat{equation}{(#3#1#4)-(#5#2#6)}
\crefmultiformat{equation}{(#2#1#3)}{ and~(#2#1#3)}{, (#2#1#3)}{ and~(#2#1#3)}

\theoremstyle{plain}
\crefname{theorem}{Theorem}{Theorems}

\crefname{corollary}{Corollary}{Corollaries}

\crefname{lemma}{Lemma}{Lemmas}

\crefname{definition}{Definition}{Definitions}

\crefname{proposition}{Proposition}{Propositions}

\crefname{appendix}{Appendix}{Appendices}

\crefname{section}{Section}{Sections}

\title{ \bf{Anisotropic drag force in finite-density QGP from charged rotating 5D black holes}}

\date{}
\author[a,b]{Sergei G. Ovchinnikov}
\affiliation[a]{Institute of Theoretical and Mathematical Physics,\\Lomonosov Moscow State University, Moscow 119991, Russia}
\affiliation[b]{HSE University, 6 Usacheva str., Moscow 119048, Russia}

\emailAdd{sovchinnikov@itmp.msu.ru}

\abstract{We study the drag force acting on a heavy quark in a holographic plasma with rotational anisotropy and finite density. The bulk dual is the CCLP black hole of five-dimensional minimal gauged supergravity, characterised by two independent rotation parameters and electric charge. In the neutral Kerr--AdS limit, we use the principal Killing string to obtain an exact drag force for arbitrary rotation parameters. The resulting force is purely tangential but generically anisotropic, reducing to the viscous form only in the equal-spin sector. We then analyse stationary strings in the charged CCLP background perturbatively in the slow-rotation regime. A regularity analysis of the Lorentzian worldsheet fixes the angular integration constants that would otherwise remain ambiguous, yielding a finite renormalised transverse drag force with a smooth Kerr--AdS limit. We also show that, in the equal-spin sector, worldsheet regularity selects a unique co-rotating equilibrium quark and compute its renormalised free-energy shift.}

\begin{document}
\maketitle
\flushbottom

\section{Introduction} 
\label{sec:introduction}

The quark-gluon plasma (QGP), created in relativistic heavy-ion collisions, is a deconfined state of strongly interacting matter \cite{Shuryak:2014zxa,Nouicer:2015jrf}. Its collective behaviour as a fluid \cite{Kovtun:2004de,Shuryak:2014zxa} indicates that, over a substantial range of temperatures relevant for phenomenology, the medium is strongly coupled, so perturbative methods of QCD become less applicable. The recent extensive experimental programme \cite{Bzdak:2019pkr,STAR:2017ckg} emphasises two important features of the QGP: finite conserved charge density and rotation. The former becomes increasingly relevant in the broader exploration of the QCD phase diagram at non-zero baryon chemical potential, particularly in the context of the NICA collider \cite{Friman:2011zz,Galatyuk:2019lcf,Senger:2021cfo} and Beam Energy Scan programme at RHIC \cite{Bzdak:2019pkr}, while the latter is motivated by the large angular momentum coming from non-central collisions and by the associated vortical structure of the plasma, which has found experimental support through global polarization observables \cite{STAR:2017ckg}.

Heavy quarks provide a particularly useful probe of the QGP. Owing to their large masses, charm and bottom quarks are produced predominantly in the earliest stages of the collision and then propagate through the evolving plasma, thereby sampling its transport properties over much of its lifetime \cite{Rapp:2018qla,Dong:2019byy,He:2022ywp}. In the regime where the momentum transfer from the medium is soft compared with the heavy-quark mass, their dynamics admits an effective description in terms of drag and momentum diffusion, or equivalently in a Langevin/Fokker-Planck language \cite{Svetitsky:1987gq,Moore:2004tg,vanHees:2004gq}. From this perspective, the drag force becomes a tractable and clear characterisation of how the medium degrades the momentum of a heavy probe.

Gauge/gravity duality offers a solid calculational framework for studying real-time transport problems at strong coupling \cite{Casalderrey-Solana:2011dxg,DeWolfe:2013cua}. Holographically, a heavy quark moving through a thermal plasma is represented by the endpoint of a trailing open string extending from the asymptotic region toward a black-hole horizon \cite{Herzog:2006gh,Gubser:2006bz}. The drag force is then identified with the flux of the conserved momentum down the trailing worldsheet. This construction, originally developed in the context of isotropic plasmas, provides a remarkably direct geometrization of dissipation and has since become one of the standard tools for analysing heavy-quark transport in strongly coupled holographic media \cite{Herzog:2006gh,Gubser:2006bz,Casalderrey-Solana:2006fio}.

This framework was later extended to charged backgrounds, including STU black holes and related finite-density settings \cite{Caceres:2006dj,Sadeghi:2009hh,Chakraborty:2014kfa,Cheng:2014fza}, showing that the presence of conserved charges can substantially modify both the drag force and the associated diffusion data. Related holographic-QCD motivated approaches have also connected heavy-quark drag to the spatial string tension \cite{Andreev:2017bvr}. More recently, Aref'eva, Golubtsova and Gourgoulhon initiated a complementary direction by considering the drag force in the five-dimensional Kerr--AdS geometry, thereby incorporating the effects of rotation and clarifying the role of the boundary frame in the holographic interpretation \cite{NataAtmaja:2010hd,Arefeva:2020jvo,Golubtsova:2021agl}.\footnote{For holographic studies of externally forced circular motion of a heavy quark, see \cite{BitaghsirFadafan:2008adl,Fadafan:2012qu}. This should be distinguished from the present setup, where rotation is encoded in the thermal state and in the bulk geometry.} From this perspective, a natural next step is to combine these two lines of development and study the drag force in a background that carries both angular momenta and electric charge. 

The purpose of the present work is to carry out this programme in the Chong–Cveti\v{c}– L\"u–Pope (CCLP) geometry \cite{Chong:2005hr}, which furnishes the general non-extremal charged rotating AdS$_5$ black hole of minimal gauged supergravity and is characterised by mass, electric charge, and two independent rotation parameters. This background provides a natural holographic arena in which one may simultaneously track the effect of a chemical potential for a conserved current and the effect of global rotation on the dynamics of a heavy probe. In this way, the CCLP solution extends the Kerr--AdS analysis by enlarging the parameter space from purely rotational deformations to a charged rotating plasma, while still preserving enough analytic control to allow a systematic trailing-string treatment and a detailed comparison with previously known neutral and non-rotating limits.

A distinctive feature of the CCLP background is that it solves not pure Einstein–Maxwell theory\footnote{In five dimensions a Kerr-Newman-AdS-like solution in pure Einstein-Maxwell theory is not known in a closed analytic form. Numerical solutions and perturbative results were found in \cite{Kunz:2007jq}.}, but five-dimensional minimal gauged supergravity, whose gauge sector necessarily contains a Chern–Simons term. In holographic terms, such bulk Chern–Simons couplings encode anomaly inflow and are tied to 't Hooft anomalies of the corresponding boundary current \cite{Kraus:2005zm,Erdmenger:2008rm}. For the purposes of the present paper, however, the central role of this structure is indirect: our drag-force calculation is governed primarily by the charged rotating geometry and by the associated worldsheet regularity conditions. Accordingly, we use the CCLP solution as a controlled holographic model of a charged rotating plasma, with the chemical potential associated with a conserved boundary current. In the same spirit, since the CCLP solution is asymptotically global AdS, the dual thermal state lives on $\mathbb R\times S^3$, and the rotation parameters have the direct interpretation of global angular velocities of this finite-volume state. At the same time, in a local-patch or high-temperature large-black-hole regime, they provide a controlled holographic proxy for the rotational anisotropy expected in strongly coupled plasma dynamics.

The paper is organised as follows. \Cref{sec:setup} sets up the holographic model, introduces the CCLP background, and reviews the analytic open-string configurations relevant for the subsequent analysis. In \cref{sec:Constant_equal-spin_solution} we study the constant equal-spin solutions and show that regularity of the Lorentzian worldsheet selects a unique co-rotating equilibrium quark; the renormalised free-energy shift of this configuration is then computed. \cref{sec:drag_force_on_the_cclp_background} is devoted to drag force. We first obtain an exact result in the neutral Kerr--AdS limit from the principal Killing string, and then analyse the charged CCLP background perturbatively in the slow-rotation regime, including the finite transverse force induced by the medium. We conclude in \cref{sec:discussion}. \cref{sec:hidden_symmetries_of_cclp_solution} summarises the hidden-symmetry structure of the CCLP solution and gives the coordinate and parameter map used in the main text, while \cref{sec:regularity_condition_for_the_worldsheet_of_the_trailing_string} contains the regularity analysis of the trailing-string worldsheet.

\section{Setup} 
\label{sec:setup}

\subsection{Holographic model of a QGP drag force}
\label{sec:setup_drag_force}

The drag force is a friction-like force induced by the medium on the heavy quark. Holographically, a heavy quark is modelled by a classical open string stretching from the asymptotic region into the bulk interior. The string is described by the Nambu--Goto action
\begin{align}
	S &= -\frac{1}{2 \pi \alpha'} \int \mathrm{d}^2 \sigma \sqrt{- g}\,,\\
	g_{\alpha \beta} &= G_{M N} \partial_\alpha X^M \partial_\beta X^N\, \hphantom{,}
\end{align}
where $g =\mathrm{det} g_{\alpha \beta}$\,, $\sigma^\alpha = (\sigma^0,\sigma^1)$ are coordinates on the string worldsheet and $X^M (\sigma)$, $M=1\ldots 5$ are the embedding functions of the string in the bulk, with $G_{M N}$ being the bulk metric. The equations of motion take the form of conserved currents, one for each coordinate:
\begin{equation}\label{eq:Nambu_Goto_equations}
	\nabla_\alpha P^\alpha_{M} = 0\,, \qquad P^{\alpha}_M =-\frac{1}{2 \pi \alpha'} G_{MN} \partial^\alpha X^ N = -\frac{1}{2 \pi \alpha' \sqrt{-g}} \,\pi^\alpha_M\,. 
\end{equation}
The $P^\alpha_M$ can therefore be identified with a current of spacetime momentum carried by the string, while $\pi^\alpha_M$ is the canonical worldsheet momentum
\begin{equation}
	\pi^\alpha_M = -(2 \pi \alpha' ) \frac{\delta S}{\delta \partial_\alpha X^M}\,.
\end{equation}
The total momentum carried by the string in the $M$-direction is then the conserved charge
\begin{equation}\label{eq:setup_momenta}
	p_M = \int \mathrm{d}\Sigma_\alpha \, P^\alpha_M =  -\frac{1}{2 \pi \alpha'} \int \mathrm{d} \sigma^1\, \pi^0_M
\end{equation}
where one integrates over a spatial cross-section of the string worldsheet, that is along the coordinate $\sigma^1$: $\mathrm{d} \Sigma_\alpha = \sqrt{-g} \,(\mathrm{d} \sigma^1,0)$. The time-independent force on the string, which can be interpreted as a drag force \cite{Gubser:2006bz,Herzog:2006gh}, is controlled by the flux through its endpoints. For a stationary configuration, the force exerted by the plasma is then given by the time derivative
\begin{equation}\label{eq:setup_drag_force_defintion}
	\frac{\partial p_M}{\partial \sigma^0} = -\frac{1}{2 \pi \alpha'} \pi^1_M\,,
\end{equation}
where we have used Nambu-Goto equations $\partial_\alpha \pi^\alpha_M = 0$\,.

Given a preferred bulk time direction $k = \partial /\partial t$, one can define the total energy of the string with respect to it by integrating over its length
\begin{equation}
	E = - \frac{1}{2 \pi \alpha'} \int \mathrm{d} \sigma^1 \,\pi_t^{0}\,.
\end{equation}

\subsection{Charged rotating black hole}
\label{sec:setup_CCLP}

We will consider 5d minimal gauged supergravity given by the action
\begin{equation}
	S = \frac{1}{2 \kappa_5^2} \int\left[ (R + 12 g^2) \star 1 - \frac{1}{2} F \wedge \star F + \frac{1}{3 \sqrt{3}} A \wedge F \wedge F \right]\,,
\end{equation}
where $A$ is the $U(1)$ graviphoton gauge field, $F= \mathrm{d} A$ is its field strength and $\Lambda = - 6 g^2$ is the cosmological constant expressed in terms of the inverse AdS radius $g = L^{-1}>0$. The Einstein and Maxwell equations of motion are
\begin{equation}
	\begin{aligned}
	R_{MN} -\frac{1}{2} F_{MP}F_N^{\hphantom{N} P} + g_{MN} \left(4 g^2 + \frac{1}{12} F_{PQ} F^{PQ}\right)=0\,,\\
	\mathrm{d} \star F = \frac{1}{3 \sqrt{3}} F \wedge F\,.
	\end{aligned}
\end{equation}

The CCLP solution is the most general known black hole solution of this theory \cite{Chong:2005hr}.\footnote{In the supersymmetric sector, a uniqueness theorem within the class of toric solutions with Calabi-type Kähler base shows that any regular black hole is locally the CCLP solution or its near-horizon geometry \cite{Lucietti:2022fqj}.} It is parameterised by four constants $a,b, m,q$, and for our purposes it is conveniently expressed in the asymptotically static Boyer-Lindquist type coordinates $x^\mu = (t,r, \theta,\phi,\psi)\,,$ where $0\leq \theta <\pi/2$, and $0\leq \phi,\psi< 2 \pi$\,.
\begin{equation}\label{eq:CCLP_solution}
	\begin{aligned}
	\mathrm{d}s^2 = 
	&- \frac{\Delta_\theta \left[(1+g^2 r^2) \rho^2 \mathrm{d} t\,+\, 2 q \nu\right]\mathrm{d} t}{\Xi_a \Xi_b \rho^2}\,+\, \frac{2 q\, \nu \omega}{\rho^2}+ \frac{f}{\rho^4} \left(\frac{\Delta_\theta \, \mathrm{d} t}{\Xi_a \Xi_b} - \omega \right)^2\,\\
	&+\, \frac{\rho^2 \mathrm{d} r^2}{\Delta_r}\,+\, \frac{\rho^2 \mathrm{d} \theta^2}{\Delta_\theta}\,+\, \frac{r^2+a^2}{\Xi_a} \sin^2 \theta \,\mathrm{d} \phi^2\,+\, \frac{r^2+b^2}{\Xi_b} \cos^2 \theta\, \mathrm{d} \psi^2\,,\\
	A &= \frac{\sqrt{3} q}{\rho^2} \left(\frac{\Delta_\theta \mathrm{d} t}{\Xi_a \Xi_b} - \omega\right)\,,
	\end{aligned}
\end{equation}
where 1-forms $\nu, \omega$ and metric functions are
\begin{equation}
	\begin{aligned}\label{eq:CCLP_metr_functions}
		&\nu = b \sin^2 \theta\, \mathrm{d} \phi\,+\, a \cos^2 \theta \,\mathrm{d} \psi\,,\qquad \omega = a \sin^2 \theta\, \dfrac{\mathrm{d} \phi}{\Xi_a} \,+\, b \cos^2 \theta \, \dfrac{\mathrm{d} \psi}{\Xi_b}\,,\\
		&\Delta_r = r^{-2} \left((r^2+a^2)(r^2+b^2)(1+ g^2 r^2) + q^2 + 2 ab q\right) - 2 m\,,\\
		&\Delta_\theta = \Xi_a \cos^2 \theta + \Xi_b \sin^2 \theta\,, \qquad \Xi_a = 1- a^2 g^2, \qquad \Xi_b = 1- b^2 g^2\,,\\
		&\rho^2 = r^2 + a^2 \cos^2 \theta + b^2 \sin^2 \theta\,, \qquad f = 2 m \rho^2 - q^2 + 2 a b q g^2 \rho^2\,,
	\end{aligned}
\end{equation}
and the rotational parameters $a,b$ are constrained such that $\Xi_a, \Xi_b \geq 0$. The horizons are located at the roots of $\Delta_r$, and we denote the largest root corresponding to the exterior horizon by $r_{+}$. The generator of the Killing horizon is
\begin{equation}\label{eq:CCLP_null_generator}
	\ell = \frac{\partial}{\partial t} + \Omega_a \frac{\partial}{\partial \phi} + \Omega_b \frac{\partial}{\partial \psi}\,,
\end{equation}
where angular velocities on the horizon are given by
\begin{equation}
	\begin{aligned}
	\Omega_a &= \dfrac{a (r_+^2 + b^2)(1+ g^2 r_+^2)\,+\, bq}{(r_+^2 + a^2)(r_+^2+ b^2)\,+\, abq}\,,\\
	\Omega_b &= \dfrac{b (r_+^2 + a^2)(1+ g^2 r_+^2)\,+\, aq}{(r_+^2 + a^2)(r_+^2+ b^2)\,+\, abq}\,.
	\end{aligned}
\end{equation}
The Hawking temperature is
\begin{equation}\label{eq:CCLP_temperature}
	T = \frac{1}{2 \pi} \dfrac{r_+^4 \left[1+ g^2 (2 r_+^2 +a^2+b^2)\right] - (a b + q)^2}{r_+ \left[ (r_+^2 +a^2) (r_+^2 + b^2) + a b q\right]}\,,
\end{equation}
and the entropy
\begin{equation}
	S = \dfrac{\pi^2 \left[ (r_+^2 +a^2) (r_+^2 + b^2) + a b q \right]}{2 \Xi_a \Xi_b r_+}\,.
\end{equation}
The CCLP solution is characterised by four independent parameters: mass, charge and two angular momenta. The angular momenta with respect to $\partial / \partial \phi$ and $\partial / \partial \psi$ are
\begin{equation}
	J_a = \dfrac{\pi \left[2 a m + q b (1+a^2 g^2)\right]}{4 \Xi_a^2 \Xi_b}\,, \qquad J_b = \dfrac{\pi \left[2 b m + q a (1+b^2 g^2)\right]}{4 \Xi_b^2 \Xi_a}\,,
\end{equation}
and the Page charge, defined as $Q = 1 /(16 \pi) \int_{S^3} (\star F - 1 / \sqrt{3} \ F \wedge A )$, is 
\begin{equation}
	Q = \dfrac{\sqrt{3} \pi q}{4 \Xi_a \Xi_b}\,.
\end{equation}
The electric potential on the horizon, defined as the contraction of the gauge field \cref{eq:CCLP_solution} with the null generator \cref{eq:CCLP_null_generator},
\begin{equation}
	\Phi_H= \ell^M A_M |_{r=r_+} =  \dfrac{\sqrt{3} q r_+^2}{(r_+^2 + a^2)(r_+^2+ b^2)\,+\, abq}\,.
\end{equation}
Finally, the holographic chemical potential is the difference in electric potentials at the asymptotic and at the horizon:
\begin{equation}\label{eq:CCLP_chem_pot}
	\mu = \Phi|_{r\rightarrow \infty} - \Phi_H  = - \Phi_H\,,
\end{equation}
where we have used that the potential vanishes at the asymptotic boundary in our gauge.

The solution has the following limits. In the neutral limit, $q=0$, one recovers Kerr--AdS, and zero rotation $a=b=0$ corresponds to Reissner-Nordstrom-AdS (RN--AdS).\footnote{Notice that Chern-Simons term trivialises for spherically symmetric configurations.} The zero-mass limit $m=0$ describes the spacetime of a point charge, and setting $q=0$ one recovers the AdS$_5$ after a diffeomorphism. We should also mention that for equal rotations\footnote{Without loss of generality, we can take $a=b$. The opposite sign is recovered by inverting the sign of $q$ and one of the angles.} $a^2=b^2$ the geometry simplifies substantially, since the $U(1)^2 \times \mathbb{R}_t$ isometry enhances to $SU(2) \times U(1) \times \mathbb{R}_t$. Finally, somewhat curiously, the CCLP solution can become a topological soliton for a certain limit of parameters \cite{Chong:2005hr,Ross:2005yj}. This is irrelevant for our purposes, however, because generic smooth horizonless configurations within the CCLP family satisfy the BPS limit \cite{Ross:2005yj,Cassani:2015upa}.

The CCLP solution is asymptotically globally AdS. The metric on the asymptotic $S^3 \times \mathbb{R}$ is recovered in the limit $r \rightarrow \infty$
\begin{equation}\label{eq:CCLP_globally_asymptotic_AdS}
	\mathrm{d} s^2|_{r \rightarrow \infty} = - g^2 \frac{\Delta_\theta r^2}{\Xi_a \Xi_b} \mathrm{d} t^2 + \frac{r^2}{\Delta_\theta} \mathrm{d} \theta^2 + \frac{r^2 \sin^2 \theta}{\Xi_a} \mathrm{d} \phi^2 + \frac{r^2 \cos^2 \theta}{\Xi_b} \mathrm{d} \psi^2 + O(r^0)\,,
\end{equation}
where $a,b$ can be removed by a diffeomorphism on $S^3$. Finally, from the absence of cross-terms at the leading order, one reads that in this chart the asymptotic is static.

For later use, let us also record the horizon structure of the RN--AdS black hole. The polynomial 
\begin{equation}\label{eq:CCLP_RN_AdS_roots}
	h(r) = r^2 \Delta_r|_{a=b=0} = g^2 r^6 + r^4 - 2 m r^2 + q^2 = g^2 (r^2 - r_+^2) (r^2 - r_{-}^2) (r^2 + r_0^2)
\end{equation}
has three pairs of roots: two real and one purely imaginary. Besides the exterior horizon $r_+$, we denote the interior horizon as $r_-$ and the remaining bit as $r_0 \geq 0$. In the limit of zero charge one obtains simple expressions for the roots
\begin{equation}\label{eq:CCLP_Swd_AdS_roots}
	r_+^2 = \frac{1}{2 g^2} \left(\sqrt{1+ 8 g^2 m} \, -1\right)\,, \qquad r_{-} = 0\,, \qquad g^2 r_{0}^2 = 1 + g^2 r_+^2\,.
\end{equation}

\subsection{Analytic open string solutions on Kerr--AdS and CCLP black holes}
\label{sec:setup_string_solutions}

Kerr--AdS geometries in all dimensions are well known for their rich structure of hidden symmetries. In particular, they admit a principal conformal Killing–Yano tensor \cite{Frolov:2017kze}, which can be used to construct principal Killing strings. These are smooth solutions of the Nambu-Goto equation describing causal open strings that stretch from the asymptotic region to the horizon, and which are regular everywhere, including the horizon \cite{Boos:2017qbx}. 

While principal Killing strings are usually expressed in the chart adapted to the null generator of the Killing horizon, we find that they take a particularly simple form in our Boyer-Lindquist type chart:
\begin{equation}\label{eq:Killing_string_Kerr_ab}
	\begin{aligned}
	\phi(t,r)   &= a g^2 t - a \int^r \mathrm{d}\bar{r} \, \dfrac{(\bar{r}^2+b^2) \Xi_a}{\bar{r}^2 \Delta_{\bar{r}}}\,, \qquad\psi(t,r) = b g^2 t - b \int^r \mathrm{d}\bar{r} \, \dfrac{(\bar{r}^2+a^2) \Xi_b}{\bar{r}^2 \Delta_{\bar{r}}}\,,\\
	\theta(t,r) &= \theta_0 = \textrm{const}\,,
	\end{aligned}
\end{equation}
where we have chosen the physical gauge $(\sigma^0,\sigma^1) = (t,r)$. This solution is stationary with respect to the natural co-rotating Killing field of the Kerr--AdS geometry. In the asymptotically static Boyer–Lindquist chart it therefore has non-trivial angular profile, and, as we show below, it carries non-vanishing axial momentum flux. Although $\partial_r\phi$ and $\partial_r\psi$ diverge at the bulk horizon, this is only a coordinate effect: in these coordinates the string winds infinitely before crossing the horizon. This is, in fact, a very generic feature of string solutions on black holes with axial symmetry and can be removed by introducing new angular variables that differ by the diverging radial term. To see this explicitly, introduce future-ingoing coordinates
\begin{equation*}
	\begin{aligned}
		v &=t \,+\, \int^r \mathrm{d}\bar r\, \frac{(\bar r^2+a^2)(\bar r^2+b^2)}{\bar r^2\Delta_{\bar r}},\\
		\phi_+ &= \phi \,+\, \int^r \mathrm{d}\bar r \,\frac{a(1+g^2\bar r^2)(\bar r^2+b^2)}{\bar r^2\Delta_{\bar r}} \,, 
		\qquad
		\psi_+ = \psi \,+\, \int^r \mathrm{d}\bar r \,\frac{b(1+g^2\bar r^2)(\bar r^2+a^2)}{\bar r^2\Delta_{\bar r}}.
	\end{aligned}
\end{equation*}
In these coordinates the principal string \cref{eq:Killing_string_Kerr_ab} takes the regular form
\begin{equation}
	\theta=\theta_0\,,\qquad \phi_+=ag^2v\,, \qquad \psi_+=bg^2v \,,
\end{equation}
and the infinite winding is only a coordinate effect.

In the vacuum Kerr--AdS case, the solution \cref{eq:Killing_string_Kerr_ab} is the natural Boyer-Lindquist representative of the principal Killing string. The equal-rotation limit $a=b$ is, however, geometrically special: the isometry group is enhanced to $SU(2) \times U(1)$ and the principal Killing structure degenerates. As a result, the equal-spin $a=b$ sector admits an additional one-parameter family of simple stationary string embeddings:
\begin{equation}\label{eq:Killing_string_constant}
	\phi(t,r) = \phi_0 + \omega t\,, \qquad\psi(t,r) =\psi_0 + \omega t\,, \qquad \theta(t,r) = \theta_0 = \textrm{const}\,,
\end{equation}
for any constant $\omega$. The $\omega=0$ case of this solution was used in the previous literature \cite{Arefeva:2020jvo,Golubtsova:2021agl} to describe a quark at rest in equilibrium with the surrounding rotating plasma. We will show below, however, that this is not correct: in a rotating plasma, equilibrium requires the dressed quark to co-rotate with the plasma. Finally, we have checked directly that these solutions exist only in the $a=b$ limit, and there are no stationary solutions with radially constant profiles for generic rotation.\footnote{The calculation is algebraically cumbersome and was carried out in a computer algebra system; a Mathematica notebook can be provided as ancillary material.}

For charged five-dimensional solutions, the situation is more complicated. The CCLP geometry does not admit an ordinary principal tensor,\footnote{There is, however, a modified object which satisfies these conditions up to a torsion 3-form given by the Hodge dual of gauge field tensor $T \propto \star F$ \cite{Kubiznak:2009qi}. Despite this, the construction of generalised principal Killing strings in this case does not appear to be known in the literature. See \cref{sec:hidden_symmetries_of_cclp_solution} for more details.} and, therefore, construction of principal Killing strings is unavailable. Unfortunately, to the best of our knowledge, no analogue of \cref{eq:Killing_string_Kerr_ab} is presently available for generic unequal rotations $a \neq b$, and we have not been able to construct one. Nevertheless, we have checked that for the equal-spin $a=b$ case, the configurations \cref{eq:Killing_string_constant} solve Nambu-Goto equations. This motivates the perturbative analysis of generic static strings carried out in \cref{sec:drag_general}.

\section{Constant equal-spin solution as equilibrium configuration}
\label{sec:Constant_equal-spin_solution}

In this section we first demonstrate that only one member of the one-parameter family \cref{eq:Killing_string_constant} is globally regular in the Lorentzian black hole geometry, and then show that this regular solution indeed describes an equilibrium heavy quark in the rotating plasma. Furthermore, we will prove the uniqueness of this configuration and discuss the consequences of non-existence of analogous configurations in unequal-spin backgrounds. Finally, we will compute the renormalised free energy of the equilibrium quark.

\subsection{Regularity and equilibrium} 
\label{sub:regularity_and_equilibrium}

Any member of the constant embeddings family \cref{eq:Killing_string_constant} experiences no drag
\begin{equation}\label{eq:no-drag-condition}
	\pi^r_M = \delta^r_M\,,
\end{equation}
and there is no momentum flow all the way to the horizon. This does not imply that any member describes an equilibrium quark: the caveat is that when the bulk rotates, these strings are regular only for a fixed value of $\omega$ due to the presence of a bulk ergoregion. In particular, for the na\"ive static solution $\omega=0$, which was used in the literature, the worldsheet metric 
\begin{equation}
	g_{\mu \nu} = G_{\mu \nu}
\end{equation}
is clearly singular at the ergosurface $G_{tt}=0$\,. A naked worldsheet singularity in the bulk exterior means that this local solution does not possess a smooth worldsheet horizon and therefore cannot be completed into a physical global saddle appropriate to equilibrium real-time physics. Equivalently, the inertial quark described by this embedding is not in a thermal state with the rotating plasma.

The induced worldsheet metric must be a regular Lorentzian geometry throughout the exterior. The form of the worldsheet metric is diagonal
\begin{equation}
	\mathrm{d}s_{ws}^2 = G(\zeta,\zeta) \mathrm{d} t^2 \,+\, \frac{\rho^2}{\Delta_r}\mathrm{d} r^2\,,
\end{equation}
where $\zeta = \partial_t+\omega(\partial_\phi+\partial_\psi)$ is the vector tangent to the constant embedding \cref{eq:Killing_string_constant}. For a physical endpoint trajectory this vector is timelike near the asymptotic boundary. Since $\Delta_r>0$ outside the bulk horizon, any zero of $G(\zeta,\zeta)$ at $r>r_+$ would make the induced worldsheet metric degenerate in the exterior. The only acceptable zero is the one located at the bulk horizon. At $r=r_+$, using
\begin{equation*}
	\zeta=\ell+(\omega-\Omega_a)(\partial_\phi+\partial_\psi),
\end{equation*}
one finds
\begin{equation}
	G(\zeta,\zeta)\big|_{r=r_+} =
(\omega-\Omega_a)^2
G(\partial_\phi+\partial_\psi,\partial_\phi+\partial_\psi)\big|_{r=r_+} \geq 0\,.
\end{equation}
The last factor is positive away from the degeneration axes of the Hopf fibre. Hence the zero of $G(\zeta,\zeta)$ is located at the bulk horizon only for
\begin{equation}\label{eq:const_string_regularity_at_horizon_corotating}
	\omega = \Omega_a\,,
\end{equation}
and $\zeta = \ell$, the horizon generator \cref{eq:CCLP_null_generator}. For this value the induced metric has a smooth worldsheet horizon coincident with
the bulk horizon.

It is then straightforward to see that this solution is regular everywhere including the bulk horizon, and, moreover, crosses it normally: there is no component tangent to a spatial cross-section of the horizon in its stationary direction. The coincidence of the two horizons also implies the equality of their Hawking temperatures. This is a good signature that the dressed quark is indeed in equilibrium with the surrounding plasma.

\subsection{Existence and uniqueness of a stationary dragless quark} 
\label{sec:existence_uniqueness_of_stationary_dragless_quark}

We see the existence of this dragless stationary configuration as a unique feature of the equal-spin geometry. More precisely, the solution \cref{eq:Killing_string_constant} is not merely a special example: within the class of stationary strings with asymptotically rigid rotation, it is \textit{the only} configuration with vanishing radial flux of spacetime momentum. To see this, consider a general stationary embedding of the form
\begin{equation}
	x^i(t,r) = x^i_0 + \omega^i t + \tilde{x}^i (r)\,,
\end{equation}
where $x^i$ collectively denotes $(\theta, \phi, \psi)$, while $x^i_0$, and $\omega^i$ are constants. For such an embedding, the absence of drag means that the radial momentum currents vanish \cref{eq:no-drag-condition}. Since we are working in a physical gauge $(\sigma^0,\sigma^1) = (t,r)$, the induced worldsheet metric is diagonal:
\begin{equation}
	0 = P^r_M\, G^{M t} = g^{r \mu} X^t_\mu = g^{tr}\,.
\end{equation}
The angular components of the momentum currents then reduce to
\begin{equation}
	\pi^r_{i} \propto g_{tt} \,\tilde{x}'^i(r)\,.
\end{equation}
Away from the worldsheet horizon $g_{tt} \neq 0$, and therefore the no-drag condition $\pi^r_i=0$ forces 
\begin{equation}
	\tilde{x}'^i(r)=0\,.
\end{equation}
Thus, $\tilde{x}^i(r)$ must be constant, so the string has no nontrivial radial profile in angular directions. In other words, within the stationary sector, vanishing drag requires the constant-profile embedding \cref{eq:Killing_string_constant}.

This observation shows that the existence of a stationary dragless quark is a special feature of the limited anisotropy of the equal-spin geometry. As we noted in \cref{sec:setup_string_solutions}, for generic unequal rotations $a \neq b$, no analogous stationary configuration exists, and, therefore, even a stationary co-rotating quark must experience a dissipative force from the plasma.

\subsection{Renormalised energy of an equilibrium heavy quark}
\label{sec:renormalised_energy_of_an_equilibrium_heavy_quark}

Having identified the unique regular stationary quark in the equal-spin plasma, we may now ask for its renormalised energy. This quantity is the natural one-body equilibrium observable associated with the dressed heavy probe, complementary to the drag force that characterises non-equilibrium momentum loss. It is also the natural analogue, in the present rotating setup, of the single-heavy-quark free-energy sector familiar from Polyakov-loop observables \cite{Kaczmarek:2002mc,Noronha:2009ud,Noronha:2010hb,Colangelo:2010pe}.

Since the boundary endpoint moves along the horizon generator $\ell$, the effective energy of the quark in its instantaneous rest frame is
\begin{equation}
	E_{eff} = - p \cdot \hat{\ell} \,,
\end{equation}
where $p$ are conserved charges defined in \cref{eq:setup_momenta}, and $\hat{\ell} = (g^2 - \Omega_a^2)^{-1/2}\, \ell$ is the unit vector along $\ell$, defined with respect to the asymptotic metric, which in BL coordinates is
\begin{equation}\label{eq:asymptotic_boundary_metric}
	\mathrm{d} s^2_{\mathbb{R} \times S^3_{\infty}} = - g^2 \mathrm{d} t^2 + \mathrm{d} \theta^2 + \sin^2 \theta \, \mathrm{d} \phi^2 + \cos^2 \theta \, \mathrm{d} \psi^2\,,
\end{equation}
Secondly, observe that the conjugate momenta satisfy
\begin{equation}
	\pi^0 \cdot \ell = 1\,,
\end{equation}
hence
\begin{equation}
	E_{eff} = + \frac{1}{2 \pi \alpha' \,\sqrt{g^2 - \Omega_a^2}} \int^{r_m}_{r_+} \mathrm{d} r\,,
\end{equation}
where we have introduced the UV cutoff $r_m$ to regularise a divergence, which in the limit $r_m \rightarrow \infty$ corresponds to an infinitely heavy quark. 

To renormalise the expression, we compare it with the energy of the quark in vacuum AdS which follows the same trajectory. The AdS limit is recovered by $T= \mu =0$, equivalently, the $m=q=0$ limit. In this case
\begin{equation}
	E_{vac}(\Omega_a) = \frac{r_m}{2 \pi \alpha'\,\sqrt{g^2 - \Omega_a^2}}\,,
\end{equation}
and the medium-induced energy shift
\begin{equation}\label{eq:const_string_thermal_correction_energy}
	\Delta E (\Omega_a; T, a, q) \equiv E_{vac}(\Omega_a) - E_{eff} (\Omega_a; T, a, q) = \frac{r_+}{2 \pi \alpha' \,\sqrt{g^2 - \Omega_a^2}}\,.
\end{equation}
This result has a simple physical interpretation. The energy shift is governed by two quantities: the horizon radius $r_+$, which sets the thermal scale of the plasma, and the co-rotation factor $\sqrt{g^2-\Omega_a^2}$, which encodes the effect of the rotating equilibrium frame. The temperature dependence is therefore inherited through $r_+$, whereas the dependence on rotation and charge is indirect, since both parameters modify the horizon radius and the horizon angular velocity. In particular, the finite-density contribution need not be monotonic.

\section{Drag force on the CCLP background} 
\label{sec:drag_force_on_the_cclp_background}

\subsection{Principal Killing string solution}
\label{sec:principal_killing_string}

The principal Killing string \cref{eq:Killing_string_Kerr_ab} provides an exact reference solution for heavy-quark drag in the neutral Kerr--AdS background, valid for arbitrary rotation parameters. The hidden-symmetry origin of this configuration is reflected in the remarkable simplicity of the worldsheet
\begin{equation}
	\mathrm{det} g_{\mu \nu} = -1\,,
\end{equation}
as well as in constant momentum fluxes
\begin{equation}\label{eq:principal_Killing_conjugate_momenta}
	\begin{aligned}
	\pi^r_\phi &= \frac{a \sin^2 \theta_0}{\Xi_a} \left[1 + g^2 \cos^2 \theta_0 (b^2-a^2)\right]\,, \qquad \pi^r_\psi   = \frac{b \cos^2 \theta_0}{\Xi_b} \left[1 + g^2 \sin^2 \theta_0 (a^2-b^2)\right]\,,\\
	\pi^r_t    &= -a g^2 \pi^r_\phi- b g^2 \pi^r_\psi\,, \qquad \pi^r_r = 1\,, \qquad \pi^r_\theta = 0\,.
	\end{aligned}
\end{equation}
Their $r$-independence allows one to determine directly the constant drag exerted on the endpoint. The vanishing of these fluxes at special values $\theta = 0, \pi/2$ is due to the degeneration of the corresponding Killing vector fields $\partial / \partial \phi$ and $\partial / \partial \psi$ on these axes.

Observe that the projections of the conjugate momenta along the directions of bulk rotation are always positive
\begin{equation}
	a \pi^r_\phi\,,\ b \pi^r_\psi \geq 0\,,
\end{equation}
which implies that the angular momentum is flowing down the string toward the bulk horizon:
\begin{equation}
	\frac{\mathrm{d} p_M}{\mathrm{d} t} = -\frac{1}{2 \pi \alpha'} \pi^r_M\,.
\end{equation}
The external force needed to keep the endpoint on that prescribed orbit is its opposite:
\begin{equation}
	F_M^{\textrm{ext}} = + \frac{1}{2 \pi \alpha'} \pi^r_M\,.
\end{equation}

Also, notice from \cref{eq:principal_Killing_conjugate_momenta} that the energy flux is not independent of the angular-momentum fluxes:
\begin{equation}
	\pi^r_t = - a g^2 \pi^r_\phi - b g^2 \pi^r_\psi\,.
\end{equation}
This implies that the power supplied by the external agent is entirely determined by the torques required to sustain the boundary rotation.

It is interesting to express the drag force in the boundary chart and connect it with the velocity of the quark. To do this, we pass from the Boyer--Lindquist angular variables of \cref{eq:CCLP_globally_asymptotic_AdS} to the round-sphere frame of \cref{eq:asymptotic_boundary_metric}. This boundary diffeomorphism rescales the two azimuthal Killing coordinates but does not mix them; correspondingly, the covector components of the flux transform in the standard way. The boundary velocity of the endpoint is
\begin{equation}
	\vv{v} = a g\, \partial_\phi + b g\, \partial_\psi\,,
\end{equation}
and the tangential drag force vector is
\begin{equation}\label{eq:principal_Killing_general_drag}
	\vv{F}_{drag} = -\frac{1}{2 \pi \alpha'} \left(\frac{\pi^r_\phi}{\sin^2 \theta_0} \partial_\phi + \frac{\pi^r_\psi}{\cos^2 \theta_0} \partial_\psi\right)\,.
\end{equation}
Notice that for $a=b$,
\begin{equation}\label{eq:principal_Killing_drag_viscous_law}
	\vv{F}_{drag} = - \frac{1}{2 \pi \alpha' \, g \Xi}\, \vv{v}\,.
\end{equation}
Thus, we see that the resulting force is generically anisotropic: for unequal rotations, the effective friction along the two azimuthal directions is different. Therefore, the drag is not, in general, collinear with the boundary velocity. Only in the equal-spin $a=b$ does the medium response reduce to the exact viscous form. At the same time, the force remains purely tangential, since the polar component vanishes identically in this exact branch.

\subsection{General stationary string on the CCLP background}
\label{sec:drag_general}

Let us now consider a general stationary string on the CCLP background. Following the prescription of \cite{NataAtmaja:2010hd}, we focus on stationary solutions whose embeddings into the asymptotically static chart \cref{eq:CCLP_solution} are time-independent up to a constant rotation
\begin{equation}\label{eq:static_drag_ansatz}
	\theta(t,r) = \theta(r)\,, \qquad \phi(t,r) = \omega_\phi \,g^2 t \,+\, \phi(r)\,, \qquad \psi(t,r) = \omega_\psi \,g^2 t \,+\, \psi(r)\,,
\end{equation}
where we have parametrised the string in physical gauge. As shown in \cref{sec:Constant_equal-spin_solution}, a non-trivial $\theta(r)$ profile is necessary for the regularity of configuration \cref{eq:static_drag_ansatz} for generic unequal rotations.

Even for static configurations, the equations of motion \cref{eq:Nambu_Goto_equations} coming from the Nambu-Goto action are too complicated to be solved in all generality. Therefore, we solve them perturbatively, by expanding in small rotation. More precisely, we rescale rotational parameters by an infinitesimal factor\footnote{Since $a,b,\omega_\varphi,\omega_\psi$ have dimensions of length, the small dimensionless quantities are the combinations $ga, gb, g \omega_\phi$ and $g \omega_\psi$. Equivalently, the expansion is a slow-rotation expansion in units of the AdS radius $L=g^{-1}$, around the RN--AdS background at fixed $m,q$ and $g$.}
\begin{equation}
	( a, b, \omega_\phi, \omega_\psi) \rightarrow \epsilon ( a, b, \omega_\phi, \omega_\psi)\,,
\end{equation}
and then expand to the first non-vanishing order in a series expansion\footnote{Substitution of \cref{eq:drag_embedding_expansions} into the full Nambu--Goto equations shows that $\phi_{(1)}$ and $\psi_{(1)}$ are fixed by the angular equations at order $\epsilon$, whereas the $\theta$-equation starts at order $\epsilon^2$ and contains no contributions from possible $\phi_{(2)}$ or $\psi_{(2)}$ terms.}
\begin{equation}\label{eq:drag_embedding_expansions}
	\begin{aligned}
	\theta(r) &=  \theta_0 \,+\, \frac{1}{2}\epsilon^2 \,\theta_{(2)}(r) \,+\, O(\epsilon^4)\,,\\
	\phi(t,r) &= \phi_0   \,+\, \epsilon\, \omega_\phi \,g^2 t \,+\, \epsilon \,\phi_{(1)}(r)  \,+\, O(\epsilon^2)\,,\\
	\psi(t,r) &= \psi_0   \,+\, \epsilon\, \omega_\psi \,g^2 t \,+\, \epsilon \,\psi_{(1)}(r)  \,+\, O(\epsilon^2)\,.
	\end{aligned}
\end{equation}
Notice that in the absence of rotation, the only regular static solution is the static string.\footnote{This can be seen by taking velocity $v$ to zero in the analysis of \cite[section 2.2.2]{NataAtmaja:2010hd}.} Also, the expansions for $\theta(r)$ must contain only even powers in $\epsilon$ by the time-reversal symmetry of the system. 

The Nambu--Goto equations give the first-order angular profiles
\begin{equation}\label{eq:drag_force_solution_phi-psi}
	\phi_{(1)}(r) = \mathfrak{p} \int_{r_+}^r \dfrac{\bar{r}^2\, \mathrm{d}\bar{r}}{h(\bar{r})} \,, \qquad 
	\psi_{(1)}(r) = \mathfrak{q} \int_{r_+}^r \dfrac{\bar{r}^2\, \mathrm{d}\bar{r}}{h(\bar{r})}\,,
\end{equation}
where $\mathfrak{p}$ and $\mathfrak{q}$ are constants, and $h(r)$ is proportional to the blackening factor of 5d RN--AdS
\begin{equation}
	h(r) = (r^{2} \Delta_r)|_{a=b=0} = r^4 (1 + g^2 r^2) - 2 m r^2 + q^2\,.
\end{equation}
For $\mathfrak{p}, \mathfrak{q} \neq 0$ these embeddings diverge on the horizon. As we have mentioned before, this is only a coordinate singularity and can be removed by introducing new angles.

The $\theta$-component of the Nambu--Goto equation first appears at order $\epsilon^2$. It gives a linear second-order ODE for $\theta_{(2)}$\,,
\begin{equation}
	\begin{aligned}
	&r h(r)\, \theta_{(2)}'' \,+\, K \theta_{(2)}' \,+\, (\mathfrak{p}^2-\mathfrak{q}^2) K_1 \,+\, (a^2-b^2)K_2 \\
	&+\, (\omega_\phi^2 - \omega_\psi^2) K_3 \,+\, (a \omega_\phi - b \omega_\psi) K_4 \,+\, (a \omega_\psi - b \omega_\phi) K_5 = 0
	\end{aligned}
\end{equation}
where $K, K_{1,2,3,4,5}$ are rational functions of $r^2$ and constants $g,m,q, \theta_0$:
\begin{equation*}
	\begin{aligned}
	K &= 2 (2 g^2 r^6 + r^4 - q^2)\,, \qquad K_1 = - \frac{r^5}{h(r)} \sin 2 \theta_0\,,\\
	K_2 &= - \frac{r}{h(r)} \left( (g^2 r^2 -1) h(r) - 2 m r^2 (1+g^2 r^2) +q^2 (2 + g^2 r^2)\right) \sin 2 \theta_0\,,\\
	K_3 &= \frac{g^4 r^9}{h(r)} \sin 2 \theta_0\,, \qquad K_4 = \frac{2 g^2 r^3 (q^2 - 2m r^2)}{h(r)} \sin 2 \theta_0\,, \qquad K_5 = \frac{2 g^2 q r^5}{h(r)} \sin 2 \theta_0\,.
	\end{aligned}
\end{equation*}
The source terms vanish in the symmetric equal-spin sector $a=b$, $\omega_\phi=\omega_\psi$, and $\mathfrak p^2=\mathfrak q^2$, as expected. For $q \neq 0$ the solution is then given as a sum over six roots of $h(r)$: the real roots $\pm r_+, \pm r_{-}$ corresponding to the outer and inner bulk horizons, and the imaginary pair $\pm i r_0$ (see \cref{eq:CCLP_RN_AdS_roots} for the discussion of the roots).
\begin{equation}\label{eq:drag_force_solution_theta}
	\begin{aligned}
	\theta_{(2)}'  &= \frac{r}{h(r)} \left\{ C r + (a^2-b^2)(g^2 r^2-1) \sin 2 \theta_0  \,-\, g^2 r^2 (\omega_\phi^2 - \omega_\psi^2) \, \sin 2 \theta_0 \vphantom{\sum_{r_* \in\, \mathcal{R}}}\right.\\
				   &+\,\left. \frac{1}{2} r \sin 2 \theta_0 \times \sum_{r_* \in\, \mathcal{R}} \left[r_*^{-1} (3 g^2 r_*^4 + 2 r_*^2 - 2m)^{-1}\,\mathcal{F}(r_*^2)\,\mathrm{log}\, \left| r-r_*\right|  \right]\right\},
	\end{aligned}
\end{equation}
where $\mathcal{R} = \{ \textrm{roots of }h(r)\}$ and $C$ is an integration constant, and 
\begin{equation}
	\begin{aligned}
		\mathcal{F}(r_*^2) 
		&=-(a^2-b^2)(2 g^2 r_*^4 + 2 (1+g^2 m)r_*^2 - 2 m - g^2 q^2)  \,-\, 2 g^2 q \,(a \omega_\psi - b \omega_\phi) r_*^2 \\
		&+\, 2 g^2 (a \omega_\phi - b \omega_\psi) \, r_*^2(1+ g^2 r_*^2)  \,-\, g^4 (\omega_\phi^2 - \omega_\psi^2) \, r_*^6 +\, r_*^2 (\mathfrak{p}^2-\mathfrak{q}^2)\,,
	\end{aligned}
\end{equation}
where we have separated the terms containing boundary velocities $\omega_{\phi, \psi}$ and angular integration constants $\mathfrak{p}, \mathfrak{q}$. The formula \cref{eq:drag_force_solution_theta} crucially depends on the number of roots of $h(r)$, and therefore does not have a direct limit for $q=0$, which will be studied separately in the next section. 

The constants $\mathfrak{p}, \mathfrak{q}$ are not arbitrary, and are, in fact, fixed by the regularity of the string worldsheet. For clarity, we relegate the regularity analysis to \cref{sec:regularity_condition_for_the_worldsheet_of_the_trailing_string} and use its results \cref{eq:app_regularity_condition_final1,eq:app_regularity_condition_final2} which express $\mathfrak{p}^2, \mathfrak{q}^2$ in terms of bulk constants and angular velocities.

We are interested in computing the drag force experienced by the fixed endpoint at the asymptotic boundary, i.e. at large $r$. It is therefore sufficient for us to extract the first few leading orders in $r$ from $\theta_{(2)}'$ rather than work with the full formula. Notice that the coefficient multiplying the logarithm in \cref{eq:drag_force_solution_theta} is odd in $r_*$, and since the contributions of the real roots $\pm r_+, \pm r_{-}$ come in pairs of opposing signs, these pairs are suppressed as 
\begin{equation}
	\mathrm{log}\, \left| r - r_{\pm} \right| - \mathrm{log}\, \left| r + r_{\pm} \right| = \mathrm{log}\, \left| \frac{r - r_{\pm}}{r + r_{\pm}}\right| = -\frac{2r_\pm}{r}+O(r^{-3})\,,
\end{equation}
while the imaginary roots $\pm i r_0$ combine into $\tan^{-1} (r/r_0) \sim O(1)$. 
\begin{equation}\label{eq:drag_force_solution_theta_asymptotic}
	\begin{aligned}
	\theta_{(2)}'  &= \frac{r^2}{h(r)} \left\{ g^2 (a^2-b^2 + \omega_\psi^2 - \omega_\phi^2)\ r \sin 2 \theta_0 \,+\, C \right.\\
				   &+\,\left. \frac{\pi}{2} \frac{\mathcal{F}(-r_0^2)\, \sin 2 \theta_0}{r_0 (3 g^2 r_0^4 - 2 r_0^2 - 2m)} \,+\, O(r^{-1}) \right\}\,.
	\end{aligned}
\end{equation}

The conjugate momenta \cref{eq:Nambu_Goto_equations}, relevant for the drag force, are given by
\begin{equation}\label{eq:drag_force_conjugate_momenta}
	\begin{aligned}
		\pi^r_\theta &= r^{-2} h(r)\, \theta_{(2)}' \epsilon^2 \,+\,O(\epsilon^3)\,,\\
		\pi^r_\phi   &= r^{-2} h(r)\, \phi_{(1)}' \,\sin^2 \theta_0 \,\epsilon \,+\, O(\epsilon^2)\,,\\
		\pi^r_\psi   &= r^{-2} h(r)\, \psi_{(1)}' \,\cos^2 \theta_0 \,\epsilon \,+\, O(\epsilon^2)\,.
	\end{aligned}
\end{equation}
The $\phi$- and $\psi$-components are regular at large $r$: using the definition \cref{eq:setup_drag_force_defintion}, angular components of the drag force are
\begin{equation}\label{eq:drag_general_angular_drag_force}
	\frac{\mathrm{d} p_\phi}{\mathrm{d} t} = -\frac{\mathfrak{p}\, \sin^2 \theta_0}{2 \pi \alpha'}  \, \,+\, O(\epsilon^2)\,, \qquad \frac{\mathrm{d} p_\psi}{\mathrm{d} t} = -\frac{\mathfrak{q}\, \cos^2 \theta_0}{2 \pi \alpha'}  \,+\, O(\epsilon^2)\,,
\end{equation}
where we have absorbed $\epsilon$ back into $\mathfrak{p},\mathfrak{q}$. The polar component of the drag force, however, diverges linearly; the coefficient of this divergence depends only on the asymptotic data $(a,b,g)$ and $\omega_{\phi,\psi}$ but not on the thermodynamic parameters. Therefore, it can be absorbed into the regularisation of the external heavy-quark source. Since for $\epsilon=0$, the bulk becomes RN--AdS and the quark \cref{eq:static_drag_ansatz} becomes static, the linear divergence is proportional to the same UV contribution that defines the rest mass of a static quark \cite{Herzog:2006gh}
\begin{equation}
	M_{rest} = \left(2 \pi \alpha'\right)^{-1} (r_m - r_+)\,,
\end{equation}
where $r_m$ is the UV cutoff. Furthermore, in \cite{Arefeva:2020jvo} it was shown that the $\theta$-component of the drag force, interpreted as a pressure gradient in the fluid, must vanish for equal-spin bulk and equal angular velocity quark, i.e. for $a=b$ and $\omega_\phi^2 = \omega_\psi^2$. Following the literature, we adopt this finite renormalisation prescription as part of our scheme. Using \cref{eq:drag_force_solution_theta,eq:drag_force_conjugate_momenta}, this prescription fixes the finite subtraction so that the integration constant $C$ drops out of the final result. In total, the renormalised transverse force at the asymptotic takes a form similar to that appearing in the previous literature \cite{Arefeva:2020jvo,Golubtsova:2021agl}
\begin{equation}
	\begin{aligned}\label{eq:drag_force_polar_final_expression}
	\left.\frac{\mathrm{d} p_\theta}{\mathrm{d} t}\right|_{r \rightarrow \infty}
	&= -g^2\,(a^2-b^2+\omega_\psi^2 - \omega_\phi^2)\, \left(M_{rest} + \frac{r_+}{2 \pi \alpha'}\right) \ \sin 2 \theta_0 \\
	&-\,\frac{1}{4 \alpha'} \frac{\mathcal{F}(-r_0^2)\, \sin 2 \theta_0}{r_0 (3 g^2 r_0^4 - 2 r_0^2 - 2m)} \,+\, O(\epsilon^3)\,,	
	\end{aligned}
\end{equation}
where we have reabsorbed $\epsilon$ back into $a,b$ and $\omega_{\phi,\psi}$, and $\mathfrak{p}^2-\mathfrak{q}^2$ are substituted via \cref{eq:app_regularity_condition_final1}.

\begin{figure}[htb]
    \centering
    \includegraphics[width=0.85\linewidth]{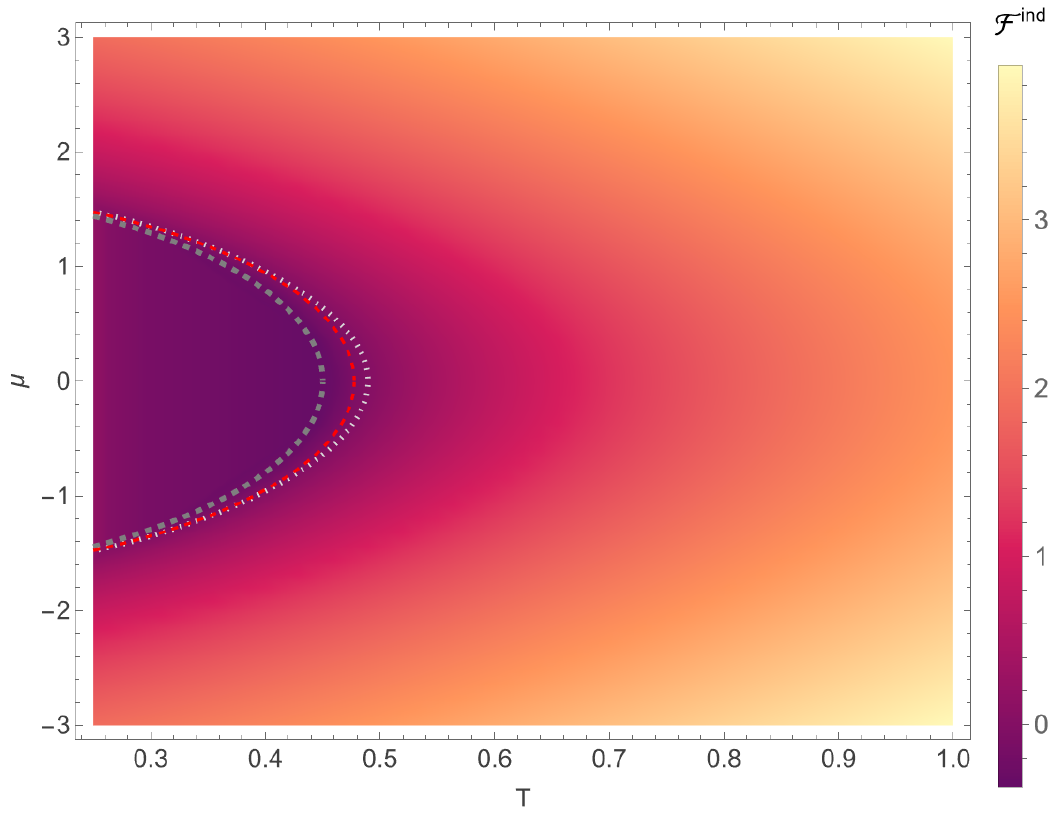}
    \caption{Heat map of $\mathcal{F}^{ind}(T,\mu)$ on the large-black-hole branch in the $(T,\mu)$ plane. The dashed curve $T_{\min}(\mu)$ represents the minimal temperature required for the black-hole branch to exist; in the region to the left of it, no black hole solution exists. The red dashed curve is the Hawking--Page curve. The white dotted curve marks $\mathcal{F}^{ind}(T,\mu) = 0$. We set $g=1$.}
    \label{fig:heatmap}
\end{figure}

\begin{figure}[ht]
	    \centering
	    \includegraphics[width=0.95\textwidth]{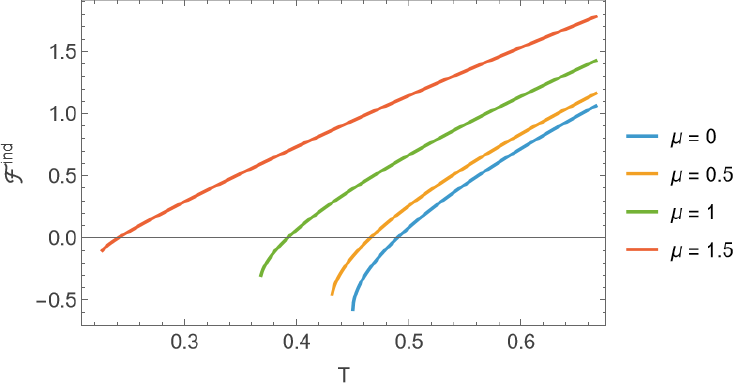}
	    \caption{Finite medium-induced transverse drag force $\mathcal{F}^{ind}(T,\mu)$ as a function of temperature for several fixed values of the chemical potential $\mu$. Negative values of $\mu$ are not shown because the drag force is an even function of $\mu$. We set $g=1$.}
	    \label{fig:fixed_chemical_potential}
\end{figure}

\begin{figure}[ht]
	    \centering
	    \includegraphics[width=0.9\textwidth]{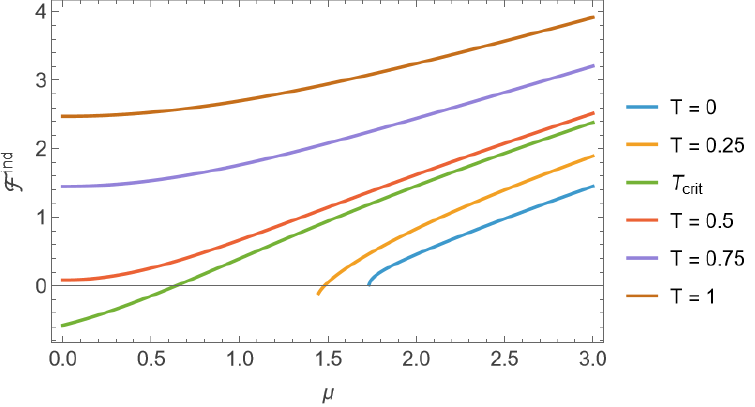}
	    \caption{Finite medium-induced transverse drag force $\mathcal{F}^{ind}(T,\mu)$ as a function of $\mu$ for several fixed values of the temperature $T$. The value $T_{\textrm{crit}} = \sqrt{2}/\pi$ is the minimal temperature for which the black hole exists for all $\mu$. We set $g=1$.}
	    \label{fig:fixed_temperature}
\end{figure}

We now study the dependence of the polar component of the drag force on thermodynamic parameters $T, \mu$. \Cref{fig:heatmap,fig:fixed_chemical_potential,fig:fixed_temperature} plot the finite medium-induced part of the transverse drag force\footnote{Notice that the overall factor $1/(2\pi\alpha')$ has been removed by the definition \cref{eq:drag_force_finite_medium_induced}.} 
\begin{equation}\label{eq:drag_force_finite_medium_induced}
	\mathcal{F}^{ind}(T,\mu) \equiv - \frac{2 \pi \alpha'}{g^2 (a^2-b^2) \sin 2 \theta_0} \left(  \left.\frac{\mathrm{d} p_\theta}{\mathrm{d}t}\right|_{\omega_{\phi,\psi} =0}  + g^2 (a^2-b^2) \sin 2 \theta_0\, M_{rest} \right)
\end{equation}
as a heat map and through one-dimensional slices at fixed chemical potential or fixed temperature. Notice from \cref{eq:drag_force_polar_final_expression} that in the absence of rotation on the boundary, the drag force does not depend on the sign of the chemical potential. This is expected since the quark is assumed to be uncharged. 

Finally, the bulk black hole does not exist for all values of $T, \mu$: from \cref{eq:CCLP_temperature,eq:CCLP_chem_pot} one can express temperature as
\begin{equation}\label{eq:drag_temperature_chem_potential_phase_diagram}
	T (r_+, \mu) = \frac{1}{2 \pi} \left( 2 g^2 r_+ + \frac{1- \mu^2 /3}{ r_+}\right)\,.
\end{equation}
For $\mu^2<3$, reality of the horizon radius $r_+$ requires the temperature to lie above the minimal-temperature curve
\begin{equation}
	T > T_{\min}(\mu) = \frac{g}{\pi} \sqrt{2 \left(1 - \frac{\mu^2}{3} \right)}\,.
\end{equation}
For $\mu^2 \geq 3$ there is one branch of black holes existing for all temperatures, which is thermodynamically dominant. For $\mu^2 < 3$ and $T>T_{\min}(\mu)$ there are two black hole branches corresponding to the roots $r_+(T,\mu)$ of \cref{eq:drag_temperature_chem_potential_phase_diagram}. The smaller-radius branch has negative heat capacity at fixed chemical potential, whereas the larger-radius branch is locally thermodynamically stable as in the standard Hawking--Page thermodynamics of charged AdS black holes \cite{Hawking:1982dh,Chamblin:1999tk}. For $\mu^2 <3$, global dominance over thermal AdS sets in only at the Hawking--Page curve
\begin{equation}\label{eq:drag_Hawking-Page_curve}
	T_{\rm HP}(\mu) = \frac{3g}{2\pi} \sqrt{1-\frac{\mu^2}{3}} = \frac{3}{2\sqrt2}\,T_{\min}(\mu)\,.
\end{equation}


\subsubsection{Kerr--AdS limit}

In the neutral limit $q=0$, the structure of the solution changes qualitatively. The reason is that the root structure of $h(r)$ degenerates, so the representation \cref{eq:drag_force_solution_theta} as a sum over roots is no longer the most convenient one. It is therefore more transparent to treat the Kerr--AdS case separately. Defining
\begin{equation}
	h_0(r) = r^{-2} h(r)|_{q=0} = g^2 r^4 +  r^2 -2m = g^2 (r^2 - r_+^2) (r^2 + r_0^2)\,,
\end{equation}
where the roots are given in \cref{eq:CCLP_Swd_AdS_roots}, one finds
\begin{equation}\label{eq:drag_force_solution_theta_zero_charge}
	\begin{aligned}
	\theta_{(2)}'  &= \frac{\sin 2 \theta_0}{h_0(r)} \left\{ g^2 r\,(a^2-b^2 - \omega_\phi^2 + \omega_\psi^2) \,+\, C \,+\, \frac{ \tan^{-1} (r/r_0)   }{g^2 r_0\, (r_+^2 + r_0^2)} \left[ (a^2-b^2) g^4 r_+^4 \right. \right.\\
				   &\left.\left. +\, g^4 r_0^4\, (\omega_\phi^2 - \omega_\psi^2) \,-\, 2 g^4 r_+^2 r_0^2 (a \omega_\phi -b \omega_\psi)  \,-\, (\mathfrak{p}^2 - \mathfrak{q}^2)\right] \,+\, \vphantom{ \frac{ \tan^{-1} (r/r_0)   }{g^2 r_0\, (r_+^2 + r_0^2)} } \mathcal{K} \, \textrm{coth}^{-1} (r/ r_+)\right\},
	\end{aligned}
\end{equation}
where $C$ and $\mathcal{K}$ are constants, and the latter depends on bulk and string data. The actual expression for $\mathcal{K}$ will not be relevant to us, because for large $r$ the term with $\textrm{coth}^{-1}$ is subleading. The inverse tangent term, however, contributes to the leading finite term.

The associated conjugate momentum is
\begin{equation}
	\pi^r_\theta = h_0(r)\, \theta_{(2)}' \epsilon^2 \,+\, O(\epsilon^3)\,,
\end{equation}

Taking a similar renormalisation scheme we recover the expression for polar drag, experienced by the endpoint at the asymptotic boundary
\begin{equation}\label{eq:drag_force_polar_final_expression_zero_charge}
	\begin{aligned}
		\left.\frac{\mathrm{d} p_\theta}{\mathrm{d} t}\right|_{r \rightarrow \infty} 
	&= -g^2\,(a^2-b^2+\omega_\psi^2 - \omega_\phi^2)\, \left(M_{rest} + \frac{r_+}{2 \pi \alpha'}\right) \ \sin 2 \theta_0 \\
	&-\, \frac{ \,\sin 2 \theta_0}{ 4 \alpha' g^2 r_0 \, (r_+^2 + r_0^2)} \times \left[ (a^2-b^2) g^4 r_+^4\,+\, g^4 r_0^4\, (\omega_\phi^2 - \omega_\psi^2) \right.\\
	&-\,\left. 2 g^4 r_+^2 r_0^2 (a \omega_\phi -b \omega_\psi) \,-\, (\mathfrak{p}^2 - \mathfrak{q}^2)\right]
	\,+\, O(\epsilon^3)\,,
	\end{aligned}
\end{equation}
where $\mathfrak{p}^2 - \mathfrak{q}^2$ is given by \cref{eq:app_regularity_condition_final1}. As a consistency check, using \cref{eq:CCLP_Swd_AdS_roots}, one can readily verify that in the limit $q \rightarrow 0$ the charged result \cref{eq:drag_force_polar_final_expression} reduces to the Kerr--AdS expression \cref{eq:drag_force_polar_final_expression_zero_charge}.

\section{Discussion} 
\label{sec:discussion}

In this paper we have studied heavy-quark probes in the charged rotating CCLP background. This allowed us to incorporate finite density and rotational anisotropy simultaneously in a single analytic holographic setup. In this sense, the present work extends the Kerr--AdS drag-force programme by adding a chemical potential through a charged rotating AdS$_5$ black hole.

The first result concerns equilibrium in a rotating plasma. In the equal-spin sector there exists a one-parameter family of constant embeddings with vanishing momentum flux, but only one of them is regular in the Lorentzian black-hole geometry. Regularity of the worldsheet selects the co-rotating string, whose horizon coincides with the bulk horizon, see \cref{eq:const_string_regularity_at_horizon_corotating}. Thus, in a rotating plasma, vanishing drag is not by itself sufficient for equilibrium: the dressed heavy quark must also define a regular thermal saddle. The corresponding renormalised energy shift \cref{eq:const_string_thermal_correction_energy} is naturally interpreted as the one-body equilibrium observable associated with this regular probe, complementary to the dissipative drag-force sector.

The second result concerns transport. In the neutral Kerr--AdS limit we used the principal Killing string to obtain an exact drag force \cref{eq:principal_Killing_general_drag} for arbitrary rotation parameters. The force is purely tangential but generically anisotropic: for unequal rotations it is not collinear with the boundary velocity, while in the equal-spin sector it reduces to the usual viscous form. In the charged CCLP background, where no exact analogue of the principal Killing string is currently known for generic unequal rotations, we analysed stationary strings perturbatively at slow rotation. The regularity condition fixes the angular integration constants entering the perturbative solution, and the resulting renormalised transverse force \cref{eq:drag_force_polar_final_expression} gives a finite-density deformation of the rotating drag problem with a smooth Kerr--AdS limit.

Several immediate extensions remain open. The most direct one is to go beyond the slow-rotation expansion and construct charged stationary strings with unequal angular momenta non-perturbatively. Such a solution would clarify whether the transverse force found here persists at finite rotation and would also determine how much of the neutral principal-string structure survives in the charged CCLP geometry. A numerical construction, similar in spirit to drag-force analyses in anisotropic finite-density backgrounds \cite{Chakraborty:2014kfa,Cheng:2014fza}, would therefore be valuable; the regularity conditions derived in \cref{sec:regularity_condition_for_the_worldsheet_of_the_trailing_string} provide natural horizon data for a shooting problem.

A second important direction is the study of fluctuations around the regular co-rotating equilibrium string. Once the correct thermal saddle has been identified, one can compute the worldsheet fluctuation spectrum, Langevin coefficients and momentum-broadening data. This would extend the equilibrium analysis of \cref{sec:Constant_equal-spin_solution} into the stochastic regime and would make direct contact with the standard holographic Brownian-motion interpretation of heavy-quark dynamics \cite{Casalderrey-Solana:2006fio,deBoer:2008gu,Son:2009vu,Giecold:2009cg}.

It would also be natural to complement the drag-force calculation by non-local heavy-quark observables in the same charged rotating background. Temporal Wilson loops and heavy-quark potentials would probe the equilibrium and screening sectors, while lightlike Wilson loops would give access to jet-quenching observables \cite{Maldacena:1998im,Rey:1998bq,Rey:1998ik,Brandhuber:1998bs,Liu:2006ug,Liu:2006he}. Spatial Wilson loops and orientation-dependent energy loss have also been studied in anisotropic finite-density holographic-QCD models \cite{Arefeva:2020bjk}, providing another useful point of comparison for future extensions of the present charged rotating setup.\footnote{For holographic-QCD treatments of heavy-quark potentials and related medium effects, see also \cite{Andreev:2006ct,Andreev:2020pqy}.} Recent rotating-plasma studies combining drag and heavy-quark potentials are especially close in spirit \cite{Chen:2023yug}. Studying these quantities together with the drag force would provide a more complete picture of heavy probes in a medium that is both rotating and at finite density.

Finally, one may ask how model-dependent the finite-density effects found here are. The CCLP gauge field belongs to five-dimensional minimal gauged supergravity and should not be identified literally with baryon number in QCD. It would therefore be useful to repeat the analysis in multi-charge gauged-supergravity backgrounds, such as the non-extremal $U(1)^3$ rotating black holes of \cite{Wu:2011gq}, where different $U(1)$ chemical potentials can be disentangled, and in bottom-up holographic-QCD constructions closer to phenomenology \cite{Arefeva:2020vae,Arefeva:2020byn,Arefeva:2018cli}. Such extensions would help separate universal consequences of rotation and finite density from features tied specifically to the CCLP embedding.

\section*{Acknowledgements}
The author is grateful to Irina Ya. Aref'eva, Oleg D. Andreev, Pavel S. Slepov, Anastasia A. Golubtsova, Marina K. Usova for the helpful discussions. SO acknowledges support by the ``BASIS" Foundation Leader Grant 24-1-1-82-5. The author is also supported by the Russian Science Foundation grant No 25-72-10177.

\appendix
\crefalias{section}{appendix}
\setcounter{equation}{0}
\renewcommand{\theequation}{\thesection.\arabic{equation}}
\numberwithin{equation}{section}

\section{Hidden symmetries of CCLP solution} 
\label{sec:hidden_symmetries_of_cclp_solution}

Introducing the chart $(t_c,x,y,\phi_c,\psi_c)$ one can recast the CCLP solution \cref{eq:CCLP_solution} in a simpler orthonormal form \cite{Kubiznak:2009qi} as
\begin{equation}
	\begin{aligned}
		&\mathrm{d} s^2 = \omega^\varepsilon \omega^\varepsilon \,+\, \sum_{i = x,y} \left(\omega^i \omega^i \,+\, \tilde{\omega}^i \tilde{\omega}^i\right)\,,\\
		&A = \sqrt{3} \left(A_P + A_Q \right)
	\end{aligned}
\end{equation}
where
\begin{equation}
	\begin{aligned}
	& \omega^x=\sqrt{\frac{x-y}{4 X}} \mathrm{d} x, \quad \tilde{\omega}^x=\frac{\sqrt{X}(\mathrm{d} t_c \,+\,y \mathrm{d} \phi_c)}{\sqrt{x(y-x)}}, \\
	& \omega^y=\sqrt{\frac{y-x}{4 Y}} \mathrm{d} y, \quad \tilde{\omega}^y=\frac{\sqrt{Y}(\mathrm{d} t_c \,+\,x \mathrm{d} \phi_c)}{\sqrt{y(x-y)}}, \\
	& \omega^{\varepsilon}=\frac{1}{\sqrt{-x y}}\left[\mu \mathrm{d} t_c \,+\,\mu(x+y) \,\mathrm{d} \phi_c \,+\, x y\, \mathrm{d} \psi_c-y A_Q-x A_P\right], \\
	& A_Q=\frac{Q}{x-y}(\mathrm{d} t_c \,+\,y \mathrm{d} \phi_c), \quad A_P = \frac{-P}{x-y}(\mathrm{d} t_c \,+\, x \mathrm{d} \phi_c).
	\end{aligned}
\end{equation}
The metric functions $X(x)$ and $Y(y)$ are given by
\begin{equation}
	\begin{aligned}
	X &= (\mu + Q)^2 + A x + C x^2 + g^2 x^3\,,\\
	Y &= (\mu + P)^2 + B y + C y^2 + g^2 y^3\,.
	\end{aligned}
\end{equation}
Constants $A, B, C, \mu, P, Q$ that parameterise the solution are not independent. There is some freedom in mapping this chart and parameterisation to the original form \cref{eq:CCLP_solution}, and, perhaps the most convenient transformation is
\begin{equation}
	x = - a^2 \cos^2 \theta - b^2 \sin^2 \theta\,, \qquad y = r^2\,,
\end{equation}
and
\begin{equation}\label{eq:app_transformation_CCLP_Kubiznak}
	\begin{aligned}
	&t 	= - t_K + (a^2 + b^2)\, \phi_K + a b \,\psi_K\,,\\
	&\phi= - a g^2\, t_K + a (1+ b^2 g^2)\, \phi_K + b \,\psi_K\,,\\
	&\psi= - b g^2\, t_K + b (1+ a^2 g^2)\, \phi_K + a \,\psi_K\,.
	\end{aligned}
\end{equation}
The parameters are expressed as
\begin{gather}
	A = b^2 + a^2 (1+b^2 g^2)\,, \qquad B = A - 2m\,, \qquad C = 1 + (a^2 + b^2) g^2\,,\\
	\mu = - a b\,, \qquad P = -q \,, \qquad Q = 0\,.
\end{gather}

The transformation \cref{eq:app_transformation_CCLP_Kubiznak} is well-defined only for $a^2 \neq b^2$\,. Then the metric possesses a non-degenerate generalised closed conformal Killing-Yano (GCCKY) 2-form \cite{Kubiznak:2009qi} given by
\begin{equation}
	h_{GCCKY} = \sqrt{-x} \, \tilde{\omega}^x \wedge \omega^x \,+\, \sqrt{-y} \tilde{\omega}^y \wedge \omega^y\,.
\end{equation}
By definition, a closed conformal Killing-Yano 2-form is a 2-form satisfying
\begin{equation}
	\nabla_M h_{NP} = \frac{1}{2} g_{M[N} (\nabla h)_{P]}\,.
\end{equation}
A generalised CCKY 2-form satisfies an analogous condition, but with the covariant derivative replaced by one with torsion. In the case of the CCLP solution, the torsion is totally antisymmetric and is given by the Hodge dual of the gauge-field tensor 
\begin{equation}
	T = \frac{1}{\sqrt{3}} \, \star F\,.
\end{equation}

The construction of the principal Killing string from CCKY is described in \cite{Boos:2017qbx}, but for GCCKY the analogous construction is not known.

\section{Regularity condition for the worldsheet of the trailing string} 
\label{sec:regularity_condition_for_the_worldsheet_of_the_trailing_string}

There are several ways to check the regularity of the string configurations, and perhaps the most straightforward way is to compute the curvature tensor of its worldsheet. Since it is two-dimensional, it is sufficient to find the scalar curvature, which evaluated on \cref{eq:drag_force_solution_theta,eq:drag_force_solution_phi-psi} schematically takes the form
\begin{equation}
	R = - 2 g^2 + \frac{4 (3 m r^2 -5 q^2)}{r^6} + \frac{L(r)}{r^8 h(r)} \epsilon^2 + O(\epsilon^4)\,,
\end{equation}
where $L(r)$ is a rational function of $r$, bulk parameters $a,b, m,q$, solution parameters $\mathfrak{p}, \mathfrak{q}$ and angular velocities $\omega_{\phi, \psi}$\,, which is too cumbersome to be written explicitly. While the leading order is regular in the exterior, the subleading order can diverge at the bulk horizon $r= r_+$, because of $h(r)$. For the solution to be regular, the numerator must therefore be a multiple of $(r-r_+)^2$\,, which is equivalent to the condition
\begin{equation}\label{eq:app_regularity_condition_full}
	\begin{aligned}
	\left( a\, \frac{q}{r_+} + b r_+ (1+g^2 r_+^2) - g^2 \omega_\psi \,r_+^3 \right)^2 - r_+^2 \mathfrak{q}^2 = \sin^2 \theta_0\, \left\{ - (a^2 - b^2) r_+^3 \left(2 \pi T + g^4 r_+^3\right) \right.\\
	\left. + 2 g^2 r_+^4 (1+ g^2 r_+^2) (a \omega_\phi -b \omega_\psi) - 2 g^2 q r_+^2 (a \omega_\psi -b \omega_\phi) - g^4 r_+^6 (\omega_\phi^2 - \omega_\psi^2) + r_+^2 (\mathfrak{p}^2 - \mathfrak{q}^2)\right\}\,.
	\end{aligned}
\end{equation}
In deriving this condition we have used \cref{eq:CCLP_RN_AdS_roots} and the temperature \cref{eq:CCLP_temperature}, expressed in terms of $r_+, r_0$,
\begin{equation}
 	T = g^2 \frac{(r_0^2 + r_+^2)(r_+^2 - r_-^2)}{2 \pi r_+^3}\,, 
\end{equation}
where $r_-^2 = r_0^2 - g^{-2}(1 + g^2 r_+^2)$\,.

The regularity condition \cref{eq:app_regularity_condition_full} in fact contains two equations. While in principle it can be solved, for example, for $\mathfrak{q}$ only, substituting the answer back into \cref{eq:drag_general_angular_drag_force} will lead to angular components of the drag force not vanishing on the corresponding axes, where this angular direction degenerates, which is unphysical. We therefore require the integration constants $\mathfrak{p}, \mathfrak{q}$ to be independent of $\theta_0$ which leads to the solution
\begin{align}\nonumber
	\mathfrak{p}^2 - \mathfrak{q}^2 &= (a^2 - b^2) r_+ \left(2 \pi T + g^4 r_+^3\right) - 2 g^2 r_+^2 (1+ g^2 r_+^2) (a \omega_\phi -b \omega_\psi) \\ \label{eq:app_regularity_condition_final1}
	  								  &+ 2 g^2 q (a \omega_\psi -b \omega_\phi) + g^4 r_+^4 (\omega_\phi^2 - \omega_\psi^2)\,,\\ \label{eq:app_regularity_condition_final2}
	\mathfrak{q}^2 &= r_+^{-2} \left( a\, \frac{q}{r_+} + b r_+ (1+g^2 r_+^2) - g^2 \omega_\psi \,r_+^3 \right)^2\,. 
\end{align}

One can then check that on the solution \cref{eq:app_regularity_condition_final1} the string is causal: $\mathrm{det} g < 0$ in the exterior.

\bibliographystyle{JHEP}
\bibliography{my_libv3}

\end{document}